\documentclass[useAMS,usenatbib,usegraphicx,psfig]{mn2e}

\title[]{Infrared spectroscopy of the remnant of Nova~Sco~2014: 
a~symbiotic star with too little circumstellar matter to decelerate the ejecta}
\author[]{U. Munari$^{1}$ $\&$  D.P.K. Banerjee$^{2}$\\
$^{1}$INAF Astronomical Observatory of Padova, 36012 Asiago (VI), Italy\\
$^{2}$Physical Research Laboratory, Navrangpura, Ahmedabad, Gujarat 380009, India\\}

\begin{document}

\maketitle

\label{firstpage}

\begin{abstract}
Pre-outburst 2MASS and WISE photometry of Nova Sco 2014 (V1534 Sco) have
suggested the presence of a cool giant at the location of the nova in the
sky.  The spectral evolution recorded for the nova did not however support a
direct partnership because no flash-ionized wind and no deceleration of the
ejecta were observed, contrary to the behavior displayed by other novae
which erupted within symbiotic binaries like V407 Cyg or RS Oph.  We have
therefore obtained an 0.8$-$2.5 $\mu$m spectra of the remnant of Nova Sco
2014 in order to ascertain if a cool giant is indeed present and if it is
physically associated with the nova.  The spectrum shows the presence of a
M6III giant, reddened by E(B-V)=1.20, displaying the typical and narrow
emission-line spectrum of a symbiotic star, including HeI 1.0830 $\mu$m with
a deep P-Cyg profile.  This makes Nova Sco 2014 a new member of the
exclusive club of novae that erupt within a symbiotic binary.  Nova Sco
2014 shows that a nova erupting witin a symbiotic binary does not always
come with a deceleration of the ejecta, contrary to the common belief.
Many other similar systems may lay hidden in past novae, expecially
in those that erupted prior to the release of the 2MASS all-sky infrared
survey, which could be profitably cross-matched now against them.
\end{abstract}

\begin{keywords}
novae, cataclysmic variables
\end{keywords}

\section{Introduction}

Classical novae (CNe) are compact binary systems, where mass is transferred
to a white dwarf (WD) - usually via the L1 Lagrange point - from a compact,
low-mass and lower-main sequence companion (see Bode \& Evans 2012 and Woudt
\& Ribeiro 2014 for recent reviews).  The electron degenerate envelope
accreted on the surface of the WD on reaching critical conditions for
ignition undergoes a thermonuclear runaway (TNR) that leads to violent mass
ejection into the surrounding empty space.  Once it is launched, the ejected
material keeps expanding at about constant velocity, with the short orbital
separation between components of the system being covered very quickly after
the TNR.  The orbital distance of CNe is of the order of $\sim$1 solar
radius and the ejecta, travelling at a typical velocity of $\sim$1000 km/s,
traverses it in the initial 10-15 minutes.  At this stage the ejecta is akin
to an extremely hot fireball which is still a few days ahead of the phase
when the nova will be discovered at optical wavelengths and/or will reach
maximum brightness.  For weeks around maximum brightness (basically until
the ejecta remains optically thick), only gradual and limited changes affect
the profile shapes and radial velocity of the emission lines 
and the P-Cyg absorptions that flank them.  Under assumptions of spherical
symmetry for the ejecta of CNe, these changes can be related to the
ionization/recombination front propagating through the ejecta as it keeps
diluting.  In a multi-component model for the ejecta (e.g., a faster bipolar
flow along the orbital axis and a slower toroidal expansion on the
equatorial plane), the picture grows more complex as considerations of the
viewing angle, actual geometrical shapes and different velocities and
densities for the components have to be taken into account.

In a minority of novae (a few \% of the total), the spectral changes around
maximum brightness can however be fairly dramatic which are manifested on a
time scale of hours to a few days.  These novae erupt within pre-existing
symbiotic binaries (hereafter termed for convenience as NwSS for "Nova
within a Symbiotic System"), where the WD orbits a late-type, cool giant
star.  Such NwSS tend to be recurrent by virtue of the high mass of the WD
and the high mass-transfer rate from the giant.  The size of the latter sets
the typical orbital period and separation between components to be $\sim$2
years and $\sim$2 AU, respectively.  It would take $\sim$100 hours for the
ejecta {\it freely} expanding at 1000 km/s to cover such a distance, so the
very large companion to the erupting WD may not be engulfed by the expanding
ejecta for some time around optical maximum.  Cool giants lose matter into
circumstellar space at a high rate through sustained stellar winds, and it
is this wind that is at the origin of the spectacular spectral changes that
affect NwSS\footnote{The NwSS should not be confused with the symbiotic
novae (SyN), which are dormant symbiotics that are turned into bright
objects by the start of stable, non-explosive, long-lasting nuclear combustion at the WD
surface (Kenyon 1986).  The non-explosive nature of the outburst means that
the lines remain sharp all the time, with the emission spectrum closely
matching that of a planetary nebula.  The SyN remain at maximum brightness
for many decades or centuries, as long as the nuclear burning continues on
the WD.  A fine example is V1016 Cyg, which turned-on half a century ago and
since then has declined in brightness by only $\sim$1 magnitude.}.  The
outburst of RS Oph in 2006 (e.g.  Evans et al.  2008, Das et al.  2006) and
V407 Cyg in 2010 (e.g.  Munari et al.  2011a) are template cases.  The
powerful initial UV-flash accompanying the TNR onset goes undetected by
observers and is lost into surrounding emptiness in the case of classical
novae whereas in the NwSS it ionizes the circumstellar gas originating from
the wind of the giant.  The resulting radiation from the recombining
flash-ionized gas is so intense that the NwSS is brought from quiescence to
maximum brightness in a very short time which is essentially the
light-travel time through the circumstellar gas.

The spectrum of the flash-ionized gas comprises of a forest of emission
lines, some emanating from atoms at high degrees of ionization, which are
quite narrow in width because the circumstellar gas originates from a low
velocity stellar wind from a giant (expansion velocity typically $\sim$ 10
km/s).  The electron density of the wind material is high enough ($n_e \sim
10^6/10^7$ cm$^{-3}$) for the recombination timescale to be of the order of
a few days.  While these narrow lines weaken, much wider lines begin to
emerge beneath them as the result of the fast expanding ejecta of the nova. 
While these latter lines grow in prominence, their FWHM however rapidly
declines as a consequence of the ejecta being slowed down while trying to
expand through the surrounding medium.  As an example, during the 2010
outburst of the NwSS V407 Cyg, Munari et al.  (2011a) measured $e$-folding
times of 4 days for the recombination of the flash-ionized gas and of 3 days
for the deceleration of the ejecta.

V1534 Sco (= Nova Sco 2014) was discovered on 2014 Mar 26.85 by K. 
Nishiyama and F.  Kabashima (CBET 3841).  Spectroscopic classification as an
He/N nova was obtained on Mar 27.8 by Ayani \& Maeno (2014), who reported a
FWHM of 7000 km/s for the H$\alpha$ line.  Joshi et al.  (2015, hereafter
J2015) studied the nova using near-IR spectroscopy covering the first 19
days of the outburst.  Their near-IR spectra confirmed an He/N
classification, with emission lines characterized by a wide trapezoidal
shape with a narrow component sitting on the top.  The positional
coincidence with a bright 2MASS cool source with an H$\alpha$ excess and the
presence of first overtone absorption bands of CO at 2.29 microns (as seen
in M type stars) led J2015 to suggest that V1534 Sco was a new member of the
exclusive NwSS club.  Initial X-ray observations obtained by Kuulkers et al. 
(2014) and Page, Osborne and Kuulkers (2014), seemed to support the NwSS
scenario, being consistent with a shock emerging from the wind of the
secondary star.  Also the accurate optical lightcurve presented by Munari et
al.  (2017) could be interpreted (not unequivocally though)
as the superposition of a recombining flashed-wind and an expanding ejecta.

There was however one  aspect seen in V1534 Sco that  conflicted with a NwSS
interpretation for it viz.  the sequence of near-IR spectra presented by
J2015 showed emission lines of near constant width which did not undergo the
rapid shrinking in velocity expected with the deceleration of the ejecta
expanding through the pre-existing wind of the M-giant companion.  In
addition to this, it may also be mentioned that no $\gamma$-ray emission was
reported for V1534 Sco, whereas such emission may have been expected to
arise from shocks associated with the high-velocity ejecta ramming in to the
wind of the M-giant, as observed in other NwSS (Ackermann et al.  2014). 
Furthermore, it will be shown below by re-analyzing some of the J2015
near-IR spectra, that no flash-ionized wind was initially present in V1534
Sco.  It was as if V1534 Sco hosted a cool giant but lacked the
extended circumstellar material that normally surrounds it.

In order to clarify its true nature, we have recently obtained a near-IR
spectrum of the remnant of V1534 Sco, aiming to answer two basic questions:
(a) is the bright 2MASS sources noted by J2015 indeed a cool giant, and if
yes (b) is it physically associated with V1534 Sco or it is just an
unrelated object that happens to lie along the same line of sight to the
nova.  Exploring these aspects will lead to a re-definition of the so far
widely accepted paradigm about a nova erupting within a symbiotic binary.

    \begin{figure*}
    \includegraphics[angle=270,width=170mm]{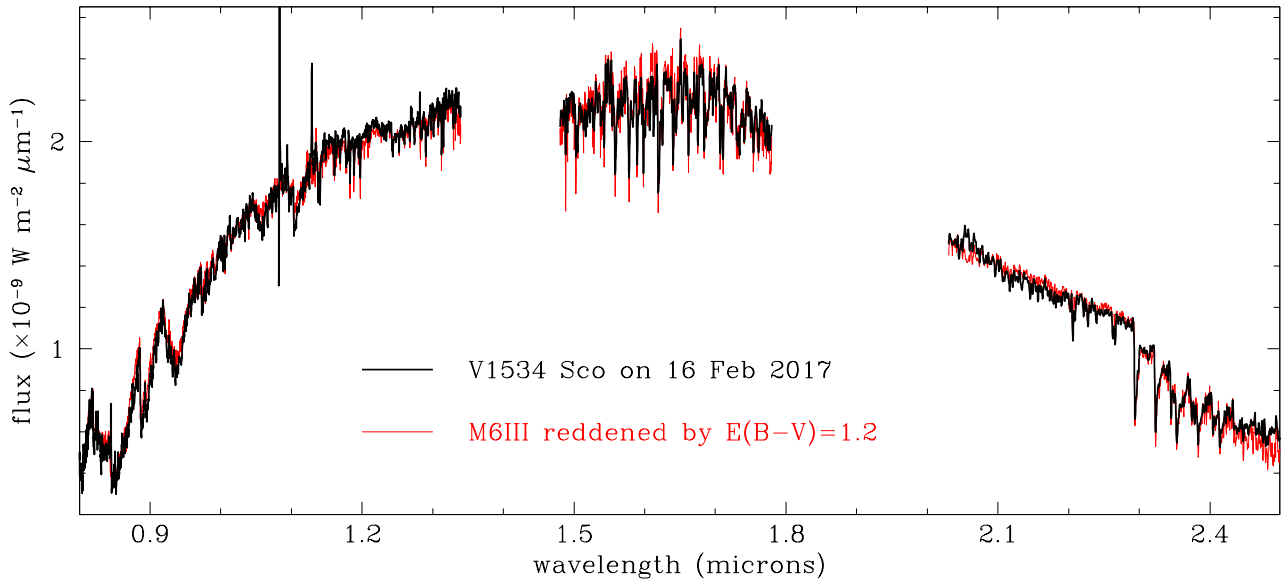}
    \caption{The IRTF-SpeX spectrum of V1534 Sco for 16 Feb 2017 is nearly
     perfectly fitted with that of an M6III template reddened by
     $E(B-V)$=1.2 following the standard $R_V$=3.1 law as tabulated by
     Fitzpatrick (1999).  The M6III template (HD 196610) is taken from the
     IRTF-SpeX spectral atlas of M stars (Cushing, Rayner, \& Vacca 2005;
     Rayner, Cushing, \& Vacca 2009).}
    \end{figure*}

\section{Observations}

V1534 Sco was observed with the SpeX spectrograph (Rayner et al.  2003) on
the 3m NASA Infra-Red Telescope Facility (IRTF), Hawaii, on 2017 February
17.63 UT.  SpeX was used in the cross-dispersed mode using a 0.5"x15" slit
resulting in a spectral coverage of the 0.77-2.50 micron region at resolving
power of 1200.  The total integration time was 319 seconds.  The A0V
star HIP 86098 was used as the telluric standard.  The data were
reduced and calibrated using the {\sc Spextool} software (Cushing et al.  2004)
and corrections for telluric absorption were performed using the IDL tool
{\sc Xtellcor} (Vacca et al.  2003).  The IRTF spectrum of V1534 Sco is presented
in Figure~1, colored in black.

\section{An M6III symbiotic binary}

\subsection{Spectral type and reddening}

The NIR spectrum of V1534 Sco in Figure~1 presents strong molecular bands,
from TiO at $\lambda$$\leq$1.0 $\mu$m and CO for $\lambda$$\geq$2.3 $\mu$m,
typical of M-type giants.  In order to derive the spectral type and
luminosity class, and measure the interstellar reddening, we have carried
out an extensive $\chi^2$ fit against the whole IRTF library for M-type
stars (Cushing, Rayner, \& Vacca 2005; Rayner, Cushing, \& Vacca 2009),
exploring the full range of reddening from $E_{B-V}$=0 to 2.0 following the
$R_v$=3.1 law as tabulated by Fitzpatrick (1999).

The best fit was found with the M6III star HD 196610, reddened by
$E_{B-V}$=1.20, which is plotted in red in Figure~1. The match is
strikingly good over the whole wavelength range and it is quite hard to
distinguish between the observed V1534 Sco spectrum and the reddened M6III
spectrum.  The first overtone $^{13}$CO absorption bands are present in
V1534 SCo along the corresponding $^{12}$CO sequence over the 2.3$-$2.5
$\mu$m spectral region, with an intensity suggesting a ratio
$^{12}$C/$^{13}$C$\sim$10.  A more accurate estimate would require a full
model atmosphere analysis (vastly outside the scope of the present paper)
which could also derive chemical abundances essential to evaluate if the
giant is on the RGB or on the AGB (eg Pavlenko et al.  2003, 2008).

In principle, while modelling the SED, a cooler M-type could be traded off
for a lower reddening, and vice-versa.  This is a major risk specifically
when only broad-band photometric data or a featureless continuum are
available for the SED modelling, as demonstrated by the different fits that
J2015 and Munari et al.  (2017) performed to available BVRI/JHK/Wise
photometry of the progenitor of V1534 Sco.  The degeneracy between intrinsic
energy distribution and reddening is removed in the present case by the
presence of molecular bands in the spectra of V1534 Sco.  Their depth is
invariant with reddening but a strong function of spectral type,
particularly for TiO (e.g.  Kenyon \& Fernandez-Castro 1987).  As
illustrated in the top panel of Figure~2, there is a good match in the depth
of TiO bands between the template M6III star and V1534 Sco.  The depth of
the TiO bands for M5III and M7III template stars would under-fit and
over-fit, respectively, those seen in V1534 Sco, no matter what value of the
the intervening reddening is considered.  Once the spectral type is fixed,
the overall shape and curvature of the spectral energy distribution returns
the reddening.  The formal errors of the $\chi^2$ fit are rather low: less
than 0.3 spectral subtypes and just 0.02 mag in the reddening.  They would
increase by incorporating the (not easily quantifiable) uncertainties
attached to the flux calibration of the IRTF spectral library and the
present spectrum of V1534 Sco, and to the Fitzpatrick (1999) reddening law.

    \begin{figure}[!Ht]
    \includegraphics[width=82mm]{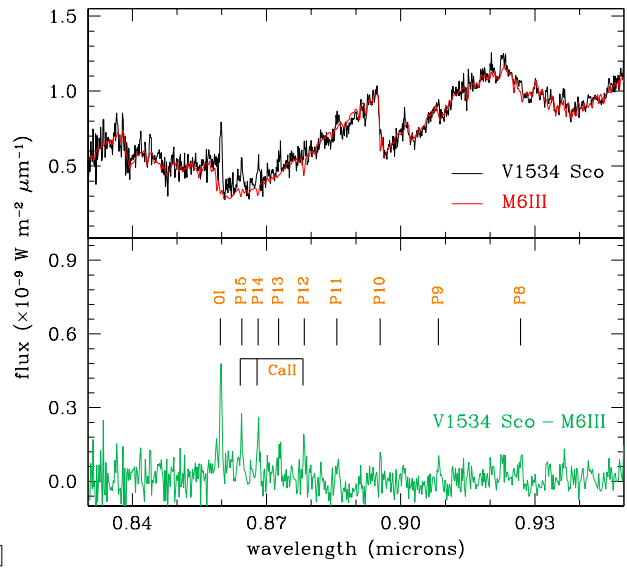}
    \caption{{\em Upper panel:} expanded view from Figure~1 around the
     far-red CaII triplet and head of the Paschen series.  {\em Lower panel}:
     a plot of the residuals after  subtracting the M6III spectrum from that of
     V1534 Sco.  The strongest emission lines are identified.}
    \end{figure}

    \begin{figure}[!Ht]
    \includegraphics[width=82mm]{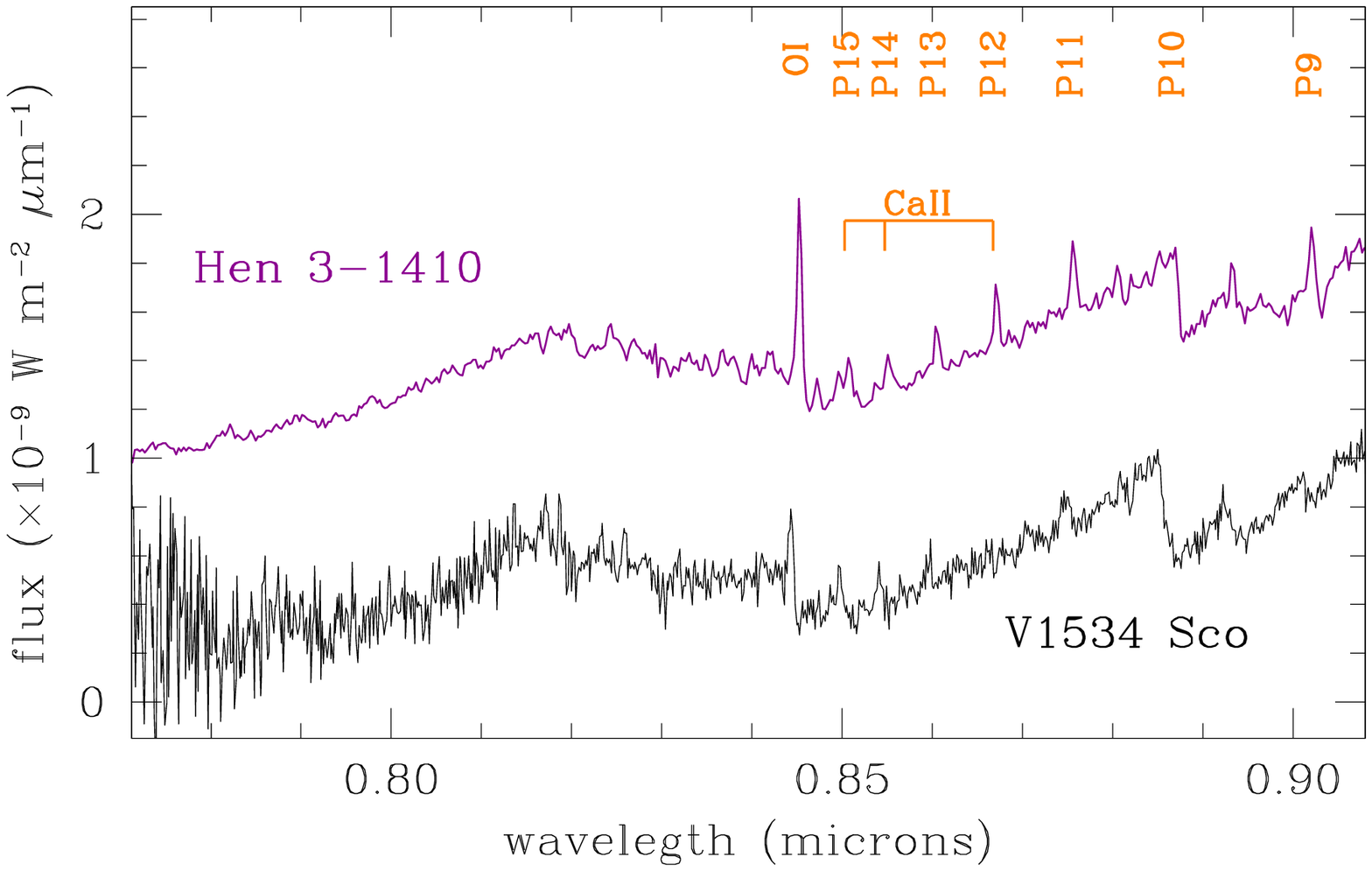}
    \caption{Comparing the blue end of the IRTF spectrum with that of the
    M6III symbiotic star Hen 3-1410 taken from the spectral atlas of
    symbiotic stars by Munari \& Zwitter (2002).}
    \end{figure}

    \begin{figure}[!Ht]
    \includegraphics[width=82mm]{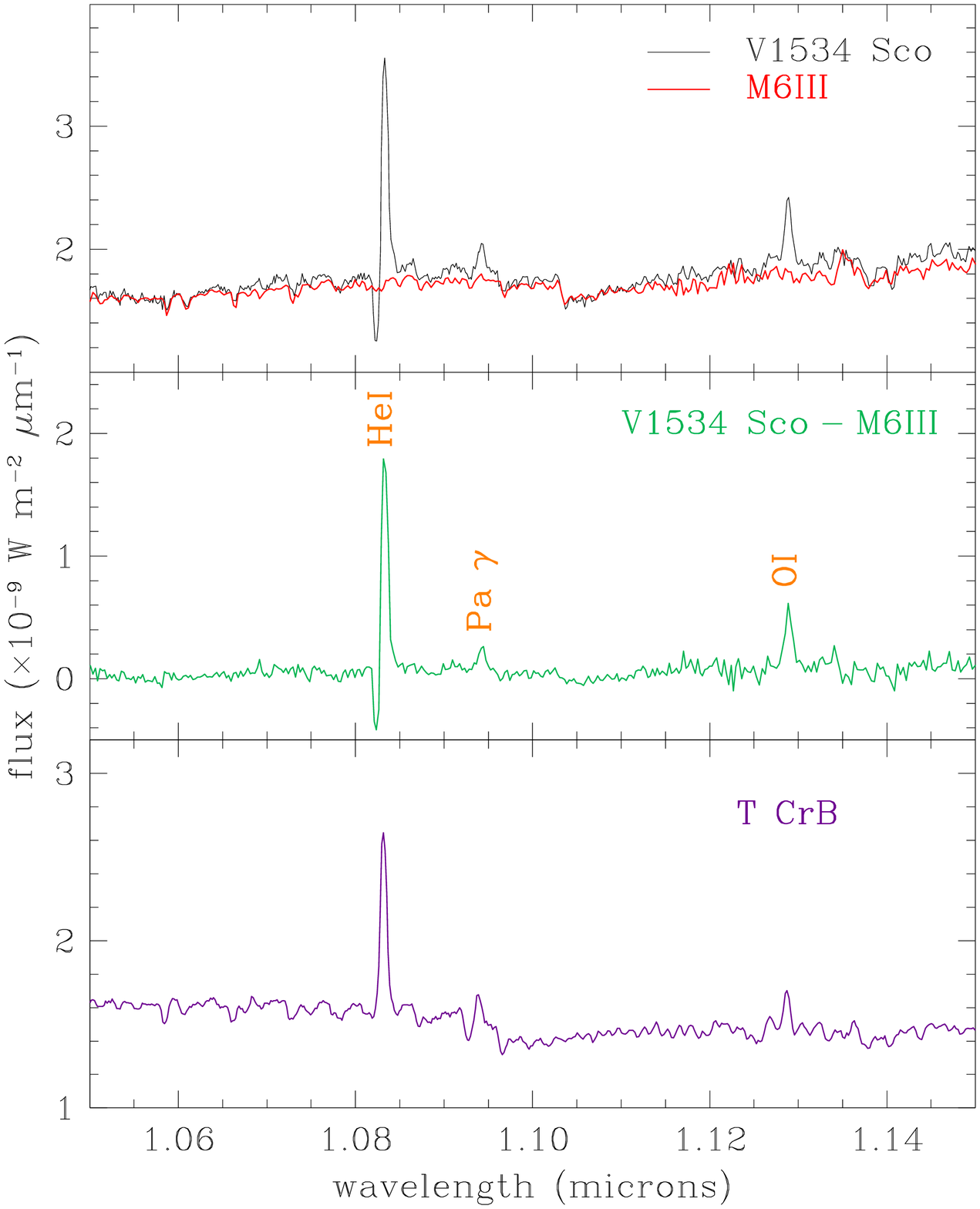}
    \caption{Similar to Figure~2, the upper panel shows an expanded view
     from Figure~1, this time around the HeI 1.0830 $\mu$m line, while the
     central panel presents the result of the subtraction of the M6III
     spectrum from that of V1534 Sco.  Note the presence of a deep P-Cyg
     absorption component for the strong HeI line.  For comparison a
     spectrum of T CrB, an M3III symbiotic star and recurrent nova, taken
     on the same night as V1534 Sco, is also shown (its flux divided
     by 100 to fit the scale).}
    \end{figure}

\subsection{The emission line spectrum}

There are several emission lines visible in the IRTF spectrum of V1534 Sco,
as illustrated in the zoomed views of Figures 2, 3, and 4.

Figure~2 focuses on the 0.84$-$0.94 $\mu$m region.  The OI 0.8446 $\mu$m
emission line is already evident in the direct spectrum of V1534 Sco of the
top panel.  Other lines like the CaII triplet and hydrogen Paschen series
become more clearly visible once the the template M6III spectrum is
subtracted from that of V1534 Sco, as illustrated in the bottom panel of
Figure~2.  We have explored the large number of $\sim$200 spectra of
symbiotic stars surveyed in the spectral atlas of Munari and Zwitter (2002),
and found that Hen 3-1410 provides the closest match with V1534 Sco.  Their
observed spectra are compared in Figure~3.  The strength of OI 0.8446 $\mu$m
and the simultaneous absence of OI 0.7772 $\mu$m is a direct evidence that a
strong Lyman-$\beta$ fluorescence (Bowen 1947) is at work in Hen 3-1410. 
This could also be the case for V1534 Sco, but the low S/N at the blue end
of its spectrum does not allow to judge about the presence and intensity of
OI~0.7772~$\mu$m.

The spectral region explored in more detail in Figure~4 is that of HeI
1.0830, Paschen-$\gamma$ 1.0938 and OI 1.1287 $\mu$m which are clearly
present in emission in the spectrum of V1534 Sco.  They well match what is
normally seen in symbiotic binaries, as proved by the comparison with T CrB
in the bottom panel of Figure~4 (this IRTF spectrum of T CrB - itself a NwSS
- was taken on the same night and with the same instrumental set-up of V1534
Sco).  The only significant difference between T CrB and V1534 Sco is the
deep P-Cyg absorption that the latter displays for HeI 1.0830 $\mu$m.  It is
not an observational artifact and there is no photospheric line at this
position that could mimic such a P-Cyg profile, as proved by the
over-plotted spectrum for the M6III template.  P-Cyg profiles are sometimes
observed for HeI lines in symbiotic stars (e.g.  Smith 1981), their presence
and intensity being usually a function of orbital phase (e.g.~Munari~1993).

The emission lines seen in V1534 Sco are all narrow and thus incompatible
with an origin in the greatly diluted ejecta of the nova.  The presence of a
strong P-Cyg absorption for HeI 1.0830 $\mu$m requires a high column
density, again incompatible with a diluted nova ejecta.  The emission lines
are therefore physically associated with the stellar system containing the
M6III giant, making it a symbiotic binary.  It is highly improbable that the
symbiotic binary and the nova are unrelated and seen just by chance along
the same line of sight.  We therefore conclude that Nova Sco 2014 erupted
from the symbiotic binary V1534 Sco, making it a new member of the exclusive
club of NwSS, which currently counts about a dozen recognized members.

    \begin{figure}
    \includegraphics[width=82mm]{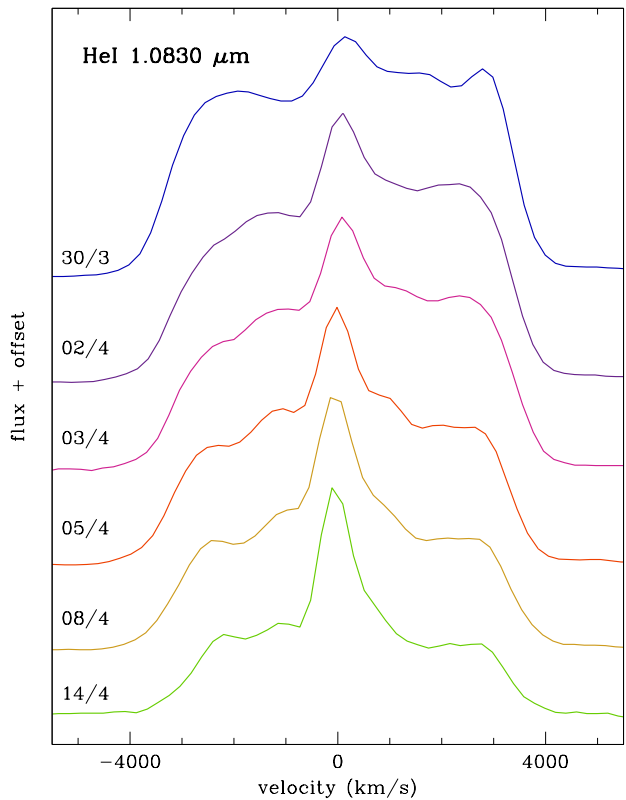}
    \caption{Evolution of the profile for HeI 1.0830 $\mu$m during the 2014 nova
    outburst of V1534 Sco, to highlight the narrow component and the broad
    pedestal. From Joshi et al. (2015) spectral data.}
    \end{figure}

\section{No flash-ionized wind or decelerating ejecta}

The absence of deceleration of the ejecta in Nova Sco 2014 was already noted
by J2015, but they did not elaborate further on this and the
narrow emission peak seen sitting on top of the broad emission lines.  We
will have a closer look to them in this section, starting from the same
near-IR spectra of Nova Sco 2014 as presented by J2015.

The profiles of the HeI 1.0830 $\mu$m emission line of Nova Sco 2014 are
shown in Figure~5, as illustrative of similar profiles affecting the other
emission lines seen in the J2015 spectra.  The two components, a broad
trapezoidal pedestal and a narrow Gaussian central peak, are quite obvious. 
We have measured from these profiles the FWHM and the integrated fluxes of
both components, with the results plotted in Figures~6 and 7.

    \begin{figure}
    \includegraphics[width=82mm]{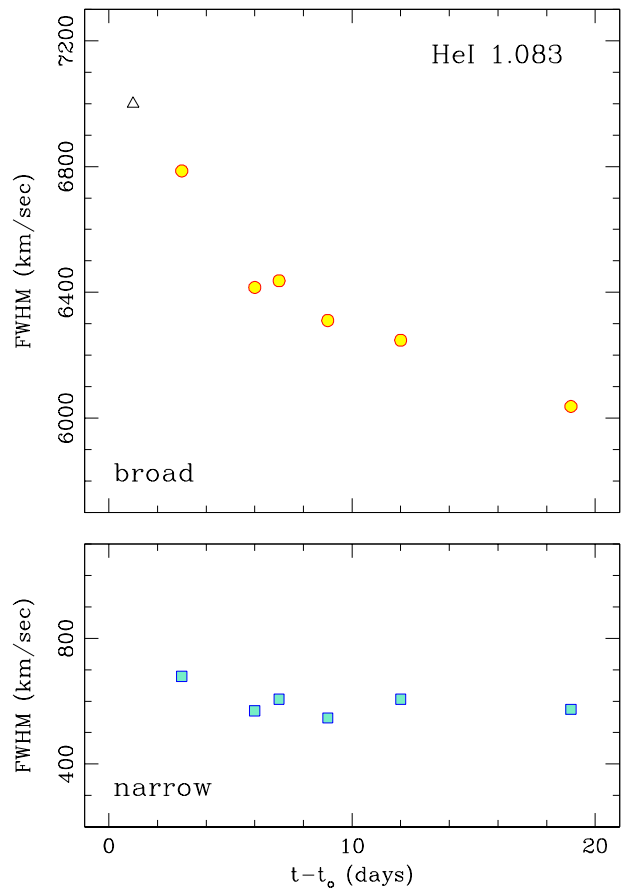}
    \caption{The evolution in width of the broad and narrow components for
    the HeI 1.0830 $\mu$m profiles in Figure 5. $t_o$ is the time of
    discovery (2014 Mar 26.85 UT). The triangle is the value measured by
    Ayani \& Maeno (2014) for H$\alpha$ on their classification spectrum.
    Note the largely expanded $y$-scale.}
    \end{figure}

    \begin{figure}
    \includegraphics[width=82mm]{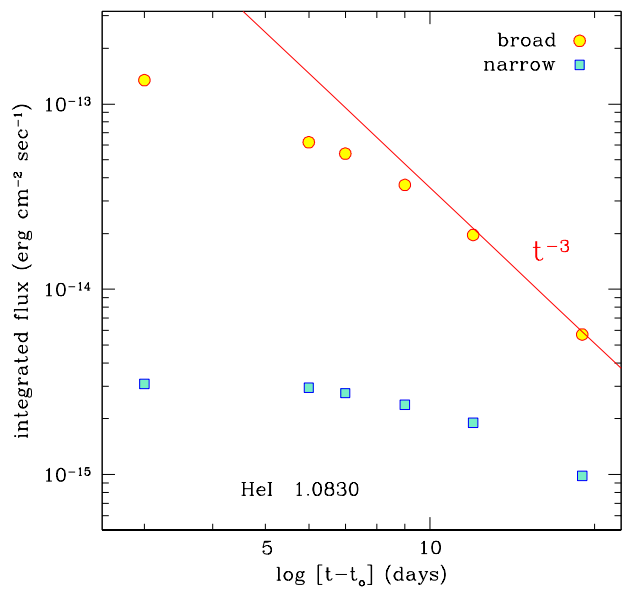}
    \caption{The evolution in integrated flux of the broad and narrow components for
    the HeI 1.0830 $\mu$m profiles in Figure 5. $t_o$ is the time of
    discovery (2014 Mar 26.85 UT).}
    \end{figure}

The broad component slowly declines in width, with an $e$-folding time of
months instead of days as for other NwSS.  This slight reduction in width is
far too slow for a deceleration caused by the ejecta ramming into any
circumstellar material.  It is instead in line with what is usually observed
in classical novae, for which ejecta freely expand into the surrounding
empty space.  This is because the recombination rate in the ejecta, and
therefore the contribution to the emission, is proportional to the local
electron density.  The latter declines quicker in the outer and faster
expanding ejecta, reducing their relative contribution to the overall line
profile which therefore appears to be narrowing (e.g.  Munari et al. 
2011b).  This interpretation is reinforced by noting in Figure~7 how the
evolution of the integrated flux of the broad component is rapidly seen
approaching the $t^{-3}$ slope of a recombining, optically thin gas.

The narrow component presents in Figure~6 a stable FWHM around $\sim$600
km/sec, or about one and half times the instrumental resolving power of
$\sim$400 km/sec in the J band.  This would prove that it cannot originate
in the quiet wind of the cool giant flash-ionized by the initial nova
TNR.  Also the evolution of its flux in Figure~7, is far removed from the
fast decline in strength associated with the rapid recombination seen in
NwSS which is completed in a matter of a few days.  Instead what is seen
here is an initial constant flux lasting about a week, then a slow decline
in line with the expectation from a thinning and diluting ejecta.  The
narrow component sitting on top of the broad trapezoidal pedestal closely
resembles the similar structural arrangement as seen in another He/N nova,
V2672 Oph (Nova Oph 2009), for which morpho-kinematical modelling (Munari et
al.  2011b) indicated in a higher density, slow-expanding equatorial torus
in the ejecta the location which produced the narrow component.  It is quite
possible that Nova Sco 2014 shares a similar bipolar morphological structure
as V2672 Oph.

From the lack of evidence for a flash-ionized wind  and also lack of
evidence for deceleration of the nova ejecta, we conclude that the M6III
giant in V1534 Sco is inefficient in engulfing its orbiting WD companion
with a wind of any significance.

\section{A revised paradigm for a nova within a symbiotic binary system}

We have been used to the notion that a nova erupting within a symbiotic
binary should inevitably face the consequences of its extended circumstellar
material viz.  a rapid and striking deceleration of the ejecta (e.g.  Munari
et al.  2011a for V407 Cyg, Das et al.  2006 for RS Oph; Banerjee et al. 
2014 for V745 Sco; Srivastava et al.  2015 for Nova Sco 2015)

V1534 Sco has proved that this is not always the case. Looking closer, we
may even find examples of transitional cases, like Nova Oph 2015 (Munari \&
Walter 2016) thereby showing a diversity characterised by the examples
below:
\begin{verse}
\vskip -0.3cm
\item[Nova Sco 2014]: no flash-ionized wind and no deceleration of the ejecta
\item[Nova Oph 2015]: modest amount of flash-ionized wind; mild
                 deceleration of the ejecta which after a short time breaks free of
                 the wind and then continues on a free expansion. 
\item[V407 Cyg]: huge amount of flash-ionized wind; prolonged and complete
                 deceleration of the nova ejecta
\end{verse}
If we leave aside any complication associated with the viewing angle in a
non-spherical morphology for the ejecta, the governing factor for
such different behavior seems to be the amount of matter shed by the cool
giant into the immediate circumstellar environment within which the WD
companion orbits.

We argue that multiple causes may lead to such differences (which may
reverse with time depending on distinct episodes of enhanced mass-loss, eg. 
Baade \& Reimers 2007), including metallicity, dust grain formation,
surface gravity, degree of Roche-lobe filling by the cool giant and amount
of irradiation by the WD companion.  They can alter by ample proportion
the fraction of mass lost to the circumstellar space via wind compared to 
that funneled through L1 directly toward the accreting~companion.

We note that other symbiotic stars may lay hidden in past novae. 
The symbiotic star in V1534 Sco would have remain undiscovered if the
spatial coincidence with a bright 2MASS source went unnoticed and if the
post-outburst observations described in this paper would not have been
performed.  No spatial coincidence with bright IR sources was normally
carried out for novae that erupted before the release of the 2MASS survey. 
Doing it now could pay some dividends and promote observations of the
remnants similar to those presented here.

\section{Acknowledgements}

We thank David Sand most warmly for making possible the IRTF observations.
We also acknowledge the useful comments by the anonymous referee.
The research at the Physical Research Laboratory is supported by the
Department of Space, Government of India.  We thank David Sand most warmly
for making possible the IRTF observations.  The Infrared Telescope Facility
is operated by the University of Hawaii under contract NNH14CK55B with the
National Aeronautics and Space Administration.

\end{document}